\documentclass[aps,prd,floatfix,reprint,amsmath,amssymb,superscriptaddress]{revtex4-2}

\usepackage{amsfonts}
\usepackage{mathrsfs}
\usepackage{amsmath}% needed for subequations
\usepackage{color}
\usepackage{natbib}
\usepackage{graphicx}
\usepackage{bm}% bold maths
\usepackage{amssymb}
\usepackage{xspace}
\usepackage{epstopdf}
\usepackage{dcolumn}% Align table columns on decimal point
\usepackage{longtable}
\usepackage{multirow}
\usepackage[colorlinks=true, letterpaper=true, pdfstartview=FitV, linkcolor=blue, citecolor=blue, urlcolor=blue]{hyperref}
\makeatother
\begin{document}

\title{Symmetry-breaking strain drives significant reduction in lattice thermal conductivity: A case study of boron arsenide}
%\title{Symmetry-breaking strain induced thermal conductivity reduction in BAs}
\author{Kaile Chen}
\affiliation{College of Physics \& Center of Quantum
Materials and Devices, Chongqing University, Chongqing 401331, China}
\affiliation{College of Physics, Chongqing Key Laboratory for Strongly Coupled Physics, Chongqing University, Chongqing 401331, P.R. China}

\author{Xin Jin}
\affiliation{College of Physics and Electronic Engineering, Chongqing Normal University, Chongqing 401331, PR China}

\author{Xiaolong Yang}
% \email[]{yangxl@cqu.edu.cn}
\email{yangxl@cqu.edu.cn}
\affiliation{College of Physics \& Center of Quantum
Materials and Devices, Chongqing University, Chongqing 401331, China}
\affiliation{College of Physics, Chongqing Key Laboratory for Strongly Coupled Physics, Chongqing University, Chongqing 401331, P.R. China}
\date{\today}% It is always \today, today,

\begin{abstract}
 Recent research has revealed that cubic boron arsenide (BAs) exhibits a non-monotonic pressure dependence of lattice thermal conductivity ($\kappa_{\rm L}$) under isotropic strain. Here, through rigorous first-principles calculations, we unveil that uniaxial tensile strain induces a monotonic reduction in the $\kappa_{\rm L}$ of BAs—a striking contrast to the isotropic scenario. The results show that applying uniaxial (100) strain leads to the lifting of phonon band degeneracy, accompanied by an overall softening of the phonon spectrum. These modifications significantly increase phonon-phonon scattering channels by facilitating the fulfillment of selection rules, resulting in a concurrent increase in both three- and four-phonon scattering rates. Consequently, $\kappa_{\rm L}$ exhibits a dramatic suppression of nearly 80\% under large tension at room temperature. Meanwhile, we unexpectedly observe that the uniaxial strain suppresses $\kappa_{\rm L}$ much more strongly in the direction perpendicular to the strain than along the stretching direction. This work establishes the fundamental understanding of the thermal conductivity behavior of BAs under uniaxial strain and opens a promising avenue for manipulating solid-state heat transport by tuning crystal symmetry.
\end{abstract}

\maketitle
\section{Introduction}
Thermal conductivity is a fundamental transport property of materials that plays a pivotal role in many applications, including heat management, thermoelectrics, memory storage devices, and thermal barrier coatings \cite{snyder2008complex,lindsay2018survey,qian2021phonon}. In non-metallic crystals, heat is primarily carried by quantized lattice vibrations, referred to as phonons \cite{ziman1960electrons}. The intrinsic thermal resistance arises from phonon-phonon scattering processes, which are highly dependent on the material's phonon dispersion, as it dictates the energy and momentum selection rules of scattering processes \cite{ravichandran2021exposing,ravichandran2020phonon}. Thus, the effective modulation of thermal transport performance can be achieved by manipulating the phonon dispersion relationship. Strain engineering provides a powerful way to tune the phonon dispersion due to its flexibility and ease of operation in experiments, which has been extensively used to modulate lattice thermal conductivity ($\kappa_{\rm L}$) in various materials \cite{tang2010lattice,JApplPhys114(2013)0649021,broido_thermal_2012,mukhopadhyay_polar_2014,parrish2014origins,lindsay2015anomalous,meng_thermal_2019,ravichandran_non-monotonic_2019,li_anomalous_2022,kundu2024electron}. 

It was long believed that the application of isotropic compressive strain (i.e., hydrostatic pressure) gives rise to a monotonic increase in $\kappa_{\rm L}$ unless a structural phase transition takes place \cite{bridgman_theoretically_1935,lindsay2015anomalous}. This can be explained as a consequence of volume shrinkage, which leads to an increase in atom density and the hardening of phonon modes \cite{ziman1960electrons}. Intriguingly, recent studies \cite{ravichandran_non-monotonic_2019,li_anomalous_2022} have revealed that some semiconductor materials with a large frequency gap between acoustic and optical branches (A-O gap), such as BAs and BSb, exhibit a non-monotonic pressure dependence of $\kappa_{\rm L}$, resulting from the competition of three- and four-phonon scattering processes. Most recently, similar anomalous behavior in $\kappa_{\rm L}$ with pressure was also reported in $\theta$-TaN \cite{kundu2021ultrahigh,kundu2024electron}, a newly discovered high-$\kappa_{\rm L}$ semimetallic material. This anomaly was found to arise from the competing effects of phonon-phonon and phonon-electron interactions under strain. These explorations have greatly advanced the understanding of strain-dependent heat conduction behavior in solids.

Despite numerous efforts towards the effect of isotropic strain or pressure on $\kappa_{\rm L}$, how the thermal conductivity responds to anisotropic strain remains largely unexplored. In fact, anisotropic strain is easier to implement in practice and also occurs in a broader range of scenarios compared to isotropic strain \cite{wang2022anomalous}. Furthermore, from a physical standpoint, the application of anisotropic strain breaks the crystal lattice symmetry, making the selection rules of phonon-phonon scattering distinct from the case of isotropic strain, which may induce unusual heat transport phenomena. In this context, exploring the effect of symmetry-breaking uniaxial strain on phonon thermal transport holds significant importance from both fundamental science and practical applications. Notably, the authors of a prior work \cite{jin2024strain} have uncovered that uniaxial strain enables the splitting of two degenerate transverse acoustic (TA) branches as well as avoided crossing between longitudinal acoustic (LA) and transverse optical (TO) branches in Wely semimetal TiO, which largely increases phonon-phonon scattering phase space, ultimately resulting in a substantial suppression of its $\kappa_{\rm L}$. This finding preliminarily underscores the essential role of uniaxial strain in tuning the solid-state thermal transport performance.  

In this work, we use cubic boron arsenide (BAs) as a showcase material to investigate the influence of uniaxial strain on phonon thermal transport properties, by utilizing first-principles based Boltzmann transport equation (BTE) with inclusion of fourth-order anharmonicity. Our calculation indicates that the application of uniaxial strain along [100] crystal direction lifts the degeneracy of transverse TA and TO branches, while softening the overall phonon spectrum. These alterations in the phonon dispersion greatly increase phonon-phonon scattering channels. As a result, the concurrent enhancement of three-phonon (3ph) and four-phonon (4ph) scattering leads $\kappa_{\rm L}$ to monotonically decrease with increasing strain. These results offer valuable insights into the response of thermal conductivity to symmetry-breaking strain.   

\section{Methodology}
In the framework of the linearized phonon BTE, the $\kappa_{\rm L}$ along the transport direction is expressed as~\cite{li_shengbte_2014}
\begin{equation}
\kappa_{\rm L}=\frac{1}{8\pi^{3}}\intop_{{\mathrm{BZ}}}\sum_{p}C_{V}(p\mathbf{q})v_{{p\mathbf{q}}}F_{{p\mathbf{q}}}d^{3}\mathbf{q},
\end{equation}
where $p$ runs over all phonon branches, BZ denotes the Brillouin zone of the crystal, and the integral is over all phonon modes from branch $p$ with wave vector $\mathbf{q}$. $\omega_{p\mathbf{q}}$, $C_V(p\mathbf{q})$, $v_{p\mathbf{q}}^\alpha$, and $F_{p\mathbf{q}}^\beta$ are the phonon angular frequency, mode heat capacity at constant volume, group velocity, and mean free displacement, respectively.
Here $F_{{p\mathbf{q}}}^{\beta}$ is calculated by exactly solving the linearized BTE starting from the relaxation-time approximation \cite{li_shengbte_2014,li2015electrical,wei2024tensile,ding2024anharmonicity}, with consideration of scattering mechanisms from 3ph, 4ph, and phonon-isotope.

According to the Fermi golden rule, the phonon scattering rates due to 3ph and 4ph processes are expressed as \cite{feng_four-phonon_2018,li_shengbte_2014,han_fourphonon_2022}
\begin{equation} \label{tau3}
    \frac{1}{\tau_{3,\lambda}}=\sum_{\lambda^{'}\lambda^{''}} \left\{\frac{1}{2}(1+{n^{0}_{\lambda^{'}}}+{n^{0}_{\lambda^{''}}}) \Gamma_- + ({n^{0}_{\lambda^{'}}}-{n^{0}_{\lambda^{''}}}) \Gamma_+ 
    \right\},
\end{equation}
\begin{equation} \label{tau4} 
 % \begin{split}
 %     \frac{1}{\tau_{4,\lambda}}=\sum_{\lambda^{'}\lambda^{''}\lambda^{'''}} \left\{\frac{1}{6}\frac{{n^{0}_{\lambda^{'}}}{n^{0}_{\lambda^{''}}}{n^{0}_{\lambda^{'''}}}}{{n^{0}_{\lambda}}} \Gamma_{--} + \frac{1}{2}\frac{(1+n^{0}_{\lambda^{'}})n^{0}_{\lambda^{''}}n^{0}_{\lambda^{'''}}}{n^{0}_{\lambda}} \Gamma_{+-} \right. \\
 %     \left. + \frac{1}{2}\frac{(1+{n^{0}_{\lambda^{'}}})(1+{n^{0}_{\lambda^{''}}}){n^{0}_{\lambda^{'''}}}}{{n^{0}_{\lambda}}}\Gamma_{++}
 %     \right\},
 % \end{split}
  \begin{split}
     \frac{1}{\tau_{4,\lambda}}=\sum_{\lambda^{'}\lambda^{''}\lambda^{'''}} &\left\{ \frac{1}{6}\frac{{n^{0}_{\lambda^{'}}}{n^{0}_{\lambda^{''}}}{n^{0}_{\lambda^{'''}}}}{{n^{0}_{\lambda}}} \Gamma_{--} \right. \\
     &\left.+ \frac{1}{2}\frac{(1+n^{0}_{\lambda^{'}})n^{0}_{\lambda^{''}}n^{0}_{\lambda^{'''}}}{n^{0}_{\lambda}} \Gamma_{+-} \right. \\
     &\left. + \frac{1}{2}\frac{(1+{n^{0}_{\lambda^{'}}})(1+{n^{0}_{\lambda^{''}}}){n^{0}_{\lambda^{'''}}}}{{n^{0}_{\lambda}}}\Gamma_{++}
     \right\},
 \end{split}
\end{equation}
in which $n^{0}_{\lambda}= \rm [exp(\hbar\omega_\lambda/\mathit{k_BT})-1]^{-1}$ is the phonon Bose-Einstein distribution at equilibrium, and $k_B$ is the Boltzmann constant. The scattering probability matrices $\Gamma_{\pm}$ and $\Gamma_{\pm\pm}$ are calculated as
\begin{equation} \label{Gamma3} 
    \Gamma_{ \pm}=\frac{\pi \hbar}{4 N}\left|V_{ \pm}^{(3)}\right|^{2} \Delta_{ \pm} \frac{\delta\left(\omega_{\lambda} \pm \omega_{\lambda^{\prime}}-\omega_{\lambda^{\prime \prime}}\right)}{\omega_{\lambda} \omega_{\lambda^{\prime}} \omega_{\lambda^{\prime \prime}}}, \\
\end{equation}
\begin{equation} \label{Gamma4} 
    \Gamma_{ \pm \pm}=\frac{\pi \hbar}{4 N} \frac{\hbar}{2 N}\left|V_{ \pm \pm}^{(4)}\right|^{2} \Delta_{ \pm \pm} \frac{\delta\left(\omega_{\lambda} \pm \omega_{\lambda^{\prime}} \pm \omega_{\lambda^{\prime \prime}}-\omega_{\lambda^{\prime \prime \prime}}\right)}{\omega_{\lambda} \omega_{\lambda^{\prime}} \omega_{\lambda^{\prime \prime}} \omega_{\lambda^{\prime \prime \prime}}}.
\end{equation}
$N$ is the number of uniformly spaced $\bf q$ points in the BZ. $\Delta$ and $\delta$ give the limitation of momentum conservation and energy conservation, respectively. In 3ph processes, $+$ represents the absorption ($\lambda_1+\lambda_2\rightarrow \lambda_3$) channels and $-$ for emission ($\lambda_1 \rightarrow \lambda_2+\lambda_3$) channels. For 4ph processes, $+ +$, $+-$ and $- -$ stand for recombination ($\lambda_1+\lambda_2+\lambda_3 \rightarrow \lambda_4$), redistribution ($\lambda_1+\lambda_2\rightarrow \lambda_3+\lambda_4$), and splitting ($\lambda_1 \rightarrow \lambda_2+\lambda_3+\lambda_4$) processes. In Eq. \eqref{Gamma3} and Eq. \eqref{Gamma4}, the scattering matrix elements $V_{ \pm}^{(3)}$ and $V_{ \pm \pm}^{(4)}$ are given by
\begin{equation} \label{v3} 
    V_{ \pm}^{(3)}=\sum_{ijk} \sum_{\alpha\beta\gamma} \Phi_{ijk}^{\alpha\beta\gamma} \frac{e_{\alpha}^{\lambda} (i) e_{\beta}^{ \pm \lambda^{\prime}} (j) e_{\gamma}^{-\lambda^{\prime \prime}} (k) } {\sqrt{\bar{M}_{i} \bar{M}_{j} \bar{M}_{k}}} e^{ \pm i \mathbf{q}^{\prime} \cdot \mathbf{r}_{j}} e^{-i \mathbf{q}^{\prime \prime} \cdot \mathbf{r}_{k}},
\end{equation}
\begin{equation} \label{v4} 
  \begin{split}
       V_{ \pm \pm}^{(4)}= \sum_{ijkl} \sum_{\alpha\beta\gamma\theta} \Phi_{ijkl}^{\alpha\beta\gamma\theta} \frac{e_{\alpha}^{\lambda} (i) e_{\beta}^{ \pm \lambda^{\prime}} (j) e_{\gamma}^{\pm \lambda^{\prime \prime}} (k) e_{\theta}^{-\lambda^{\prime \prime \prime}} (l) } {\sqrt{\bar{M}_{i} \bar{M}_{j} \bar{M}_{k} \bar{M}_{l}}} \\
       \times e^{ \pm i \mathbf{q}^{\prime} \cdot \mathbf{r}_{j}} e^{ \pm i \mathbf{q}^{\prime \prime} \cdot \mathbf{r}_{k}} e^{-i \mathbf{q}^{\prime \prime \prime} \cdot \mathbf{r}_{l}},
  \end{split}
\end{equation}
where $\Phi_{ijk}^{\alpha\beta\gamma}$ and $\Phi_{ijkl}^{\alpha\beta\gamma\theta}$ are the third-order and fourth-order interatomic force constants (IFCs). The atoms are signed by \textit{i, j, k, l}, and directions are differed by $\alpha, \beta, \gamma, \theta$. $e_{\alpha}^{\lambda}$ represents the phonon eigenvector component and $\mathbf{r}_j$ is the position vector of the $\textit{j}$th unit cell. It can be seen from Eqs.~\eqref{v3} and \eqref{v4} that the scattering matrix elements are determined by the anharmonic IFCs as well as the spatial overlap of the corresponding modes.

The weighted phase space (WPS) due to 3ph and 4ph scattering is given by 
\begin{equation} \label{W3} 
 \begin{split}
 W_{\pm}=\frac{1}{2N}\sum_{\lambda^{\prime}\lambda^{\prime\prime}} 
 \begin{Bmatrix}
 2\left(n_{ \lambda^ \prime}^0 - n_{ \lambda^{\prime \prime}}^0 \right)\\
 n_{ \lambda^ \prime}^0 + n_{ \lambda^{\prime \prime}}^0+1
 \end{Bmatrix}
 \frac{\delta \left(\omega_{\lambda}\pm\omega_{\lambda\prime}-\omega_{\lambda^{\prime\prime}} \right)}{\omega_\lambda{\omega_{\lambda\prime}}{\omega_{\lambda^{\prime\prime}}}},
 \end{split}
\end{equation} 

\begin{equation} \label{W4} 
 \begin{split}
 W_{\pm \pm}=\frac{1}{6N^2}\sum_{\lambda^{\prime} \lambda^{\prime\prime} \lambda^{\prime\prime\prime}} 
 \begin{Bmatrix}
 3\left(1+ n_{ \lambda^ \prime}^0  \right) \left(1+ n_{ \lambda^ \prime\prime}^0 \right) n_{ \lambda^ \prime\prime\prime}^0 \\
 3\left(1+ n_{ \lambda^ \prime}^0  \right)  n_{ \lambda^ \prime\prime}^0  n_{ \lambda^ \prime\prime\prime}^0 \\
 n_{ \lambda^ \prime}^0 n_{ \lambda^ \prime\prime}^0 n_{ \lambda^ \prime\prime\prime}^0
 \end{Bmatrix} \\
 \times \frac{\delta \left(\omega_{\lambda} \pm \omega_{\lambda\prime} \pm \omega_{\lambda^{\prime\prime}} -\omega_{\lambda^{\prime\prime\prime}} \right)} {\omega_\lambda {\omega_{\lambda\prime}} {\omega_{\lambda^{\prime\prime}}} {\omega_{\lambda^{\prime\prime\prime}}}}.
 \end{split}
\end{equation}
These quantities can characterize the number of all available phonon scattering channels for simultaneously satisfying the energy and momentum conservation.

All first-principles calculations were performed using the projector-augmented plane wave method \cite{blochl_projector_1994} as implemented in the Vienna Ab initio Simulation Package \cite{kresse1993ab,kresse1996efficient}. The exchange-correlation functional was adopted using the generalized gradient approximation of the Perdew-Burke-Ernzerhof functional \cite{perdew2008restoring,le2012physisorption}. A plane-wave energy cut-off of 520 eV was used for all the calculations. The harmonic and anharmonic (third- and fourth-order) IFCs were calculated by a finite displacement approach in a $4\times4\times4$ supercell with a $3\times3\times3$ $\mathbf{q}$ mesh, using Phonopy \cite{togo2015first}, Thirdorder \cite{li_shengbte_2014}, and Fourthorder \cite{han_fourphonon_2022} codes, respectively. The interaction radii were truncated to 0.5 nm and 0.35 nm for third-order IFCs and fourth-order IFCs, respectively. With these IFCs, the lattice thermal conductivity was obtained by iteratively solving the BTE through the FourPhonon package \cite{li_shengbte_2014,han_fourphonon_2022} on a $16\times16\times16$ \textbf{q}-mesh grid, which has been proven to afford the convergence of $\kappa_{\rm L}$ \cite{feng_four-phonon_2017,yang_stronger_2019}.

\section{Results and discussion}
\subsection{Monotonic decrease of thermal conductivity under uniaxial strain}
\begin{figure}[htp]
\centering   
\includegraphics[width=1.0\columnwidth]{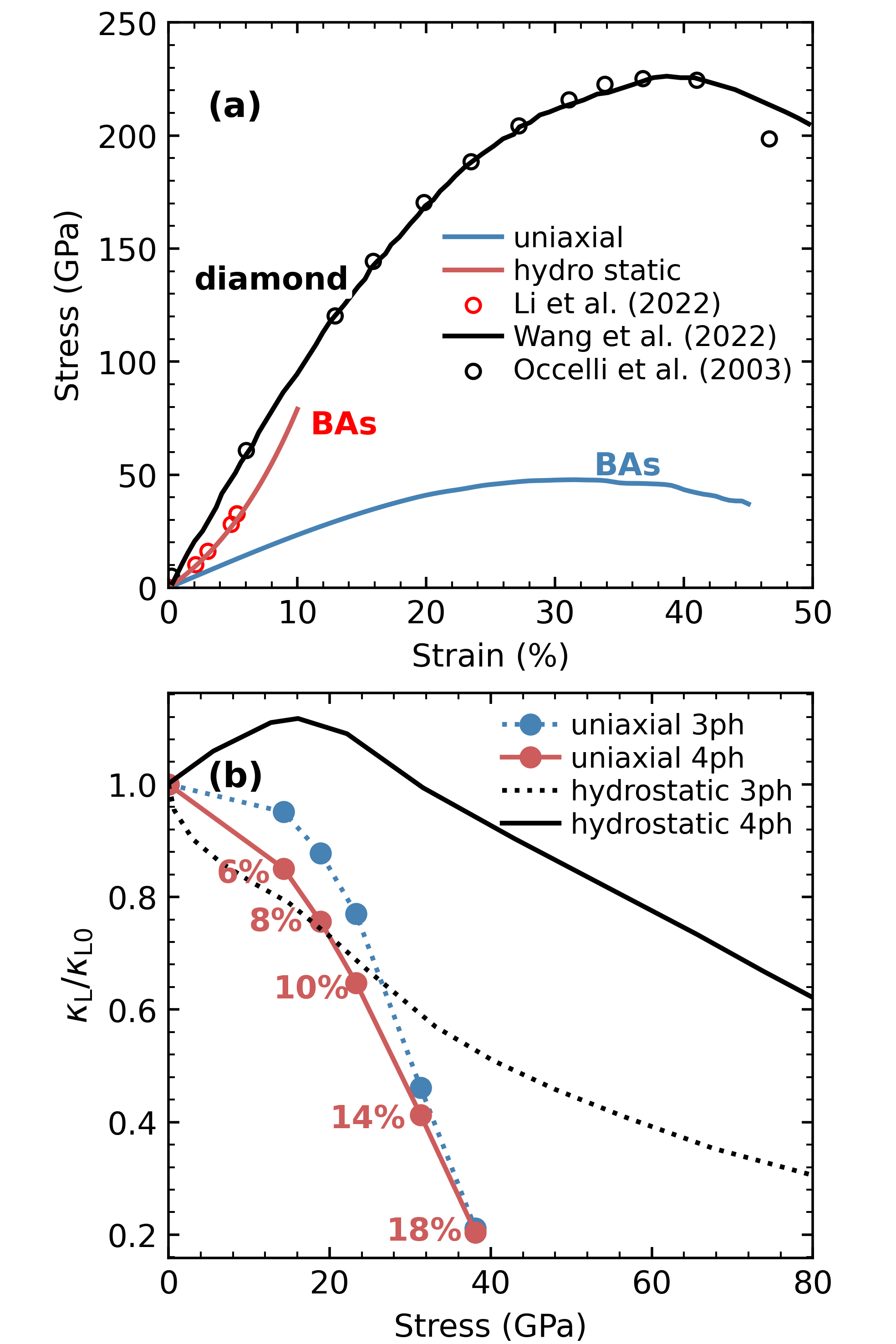}
\caption{(a) Stress-strain curves of BAs under uniaxial tension and hydrostatic pressure, compared to the literature data for BAs and diamond \cite{li_anomalous_2022,telling_theoretical_2000,wang_anomalous_2022}. (b) Stress-dependent $b$-axis $\kappa_{\rm L}$ of BAs in comparison with available results under hydrostatic pressure, calculated with (solids lines) and without (dashed lines) including 4ph scattering.}
\label{Fig1}
\end{figure}

We begin by examining the mechanical properties of BAs under uniaxial tensile strain along the [100] crystallographic direction. Fig.~\ref{Fig1}(a) shows the stress-strain response of BAs under uniaxial tension, along with corresponding data for diamond from the literature \cite{li_anomalous_2022,telling_theoretical_2000,wang_anomalous_2022}. The hydrostatic pressure dependence of BAs on compressive strain is also provided for comparison. For diamond, prior first-principles calculations yield the results in good agreement with experimental measurements, validating the computational methodology. Applying the same approach, we obtain the stress-strain curve of BAs under uniaxial tensile loading. It is clearly seen from Fig.~\ref{Fig1}(a) that the maxima in the stress, i.e., the ultimate tensile strength, occurs at $\sim$30\% strain for BAs, significantly lower than diamond that reaches its stress peak at $\sim$40\% strain. Remarkably, BAs exhibits a nearly linear stress-strain response under uniaxial strains below $<$20\%, indicative of elastic deformation behavior within this strain regime. On this basis, in the present work, we focus our investigation on the thermal transport properties of BAs at $\varepsilon <$ 20\%. Besides, we observe that the hydrostatic pressure in strained BAs is substantially higher than its tensile strength. This notable difference suggests that imposing uniaxial strain in BAs is experimentally more feasible than hydrostatic pressure loading.

Figure~\ref{Fig1}(b) presents the calculated stress dependence of RT $\kappa_{\rm L}$ for BAs along the direction perpendicular to the uniaxial loading direction (corresponding to $b$ axis). It is apparent that the $b$-axis $\kappa_{\rm L}$ decreases monotonically with increasing stress, regardless of whether 4ph scattering is considered. Most strikingly, the $\kappa_{\rm L}$ undergoes an abrupt decline when the strain exceeds 6\%, e.g., nearly 60\% reduction observed at $\varepsilon =$ 14\%. This behavior sharply contrasts with the behavior observed under hydrostatic pressure \cite{ravichandran_non-monotonic_2019,li_anomalous_2022}, where $\kappa_{\rm L}$ initially increases and then decreases, as seen in Fig.~\ref{Fig1}(b). In the following sections, we explore the underlying mechanisms for the monotonic decrease of $\kappa_{\rm L}$ under uniaxial strain in detail.

\subsection{Uniaxial strain-induced phonon degeneracy lifting and enlarged scattering phase space}
\begin{figure}[htp]
\centering
\includegraphics[width=1.05\columnwidth]{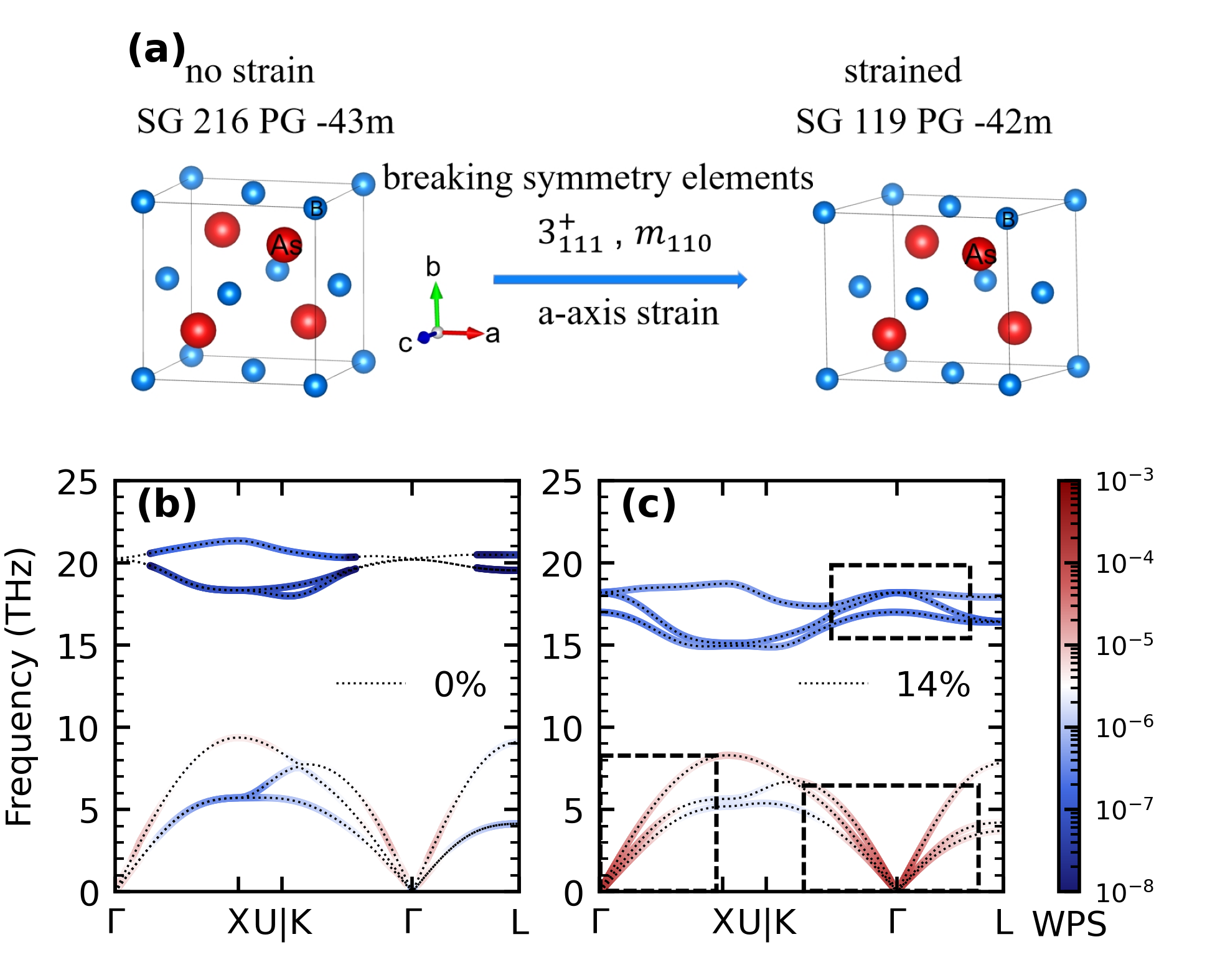}
\caption{(a) Schematic diagram of crystal symmetry breaking induced by uniaxial strain along the [100] direction. Phonon dispersions and projected 3ph phase space of BAs under 0\% (b) and 14\% (c) tensile strain.}
\label{Fig2}
\end{figure}

BAs is a recently discovered semiconductor material with high thermal conductivity exceeding 1000 W/mK at RT \cite{kang_experimental_2018,li_high_2018,tian_unusual_2018}. This ultrahigh $\kappa_{\rm L}$ is largely due to its unique phonon band structure, including a large A-O gap and a bunching of acoustic branches. These features strongly impede scattering processes involving three acoustic phonons (AAA) and two acoustic phonons combining into one optical phonon (AAO) \cite{lindsay_first-principles_2013}, leading to unusually weak 3ph scattering. While the phonon dispersion of BAs largely prevents the 3ph processes, it does not constrain 4ph processes, thereby enabling 4ph scattering to play a significant role in lowering $\kappa_{\rm L}$ \cite{feng_four-phonon_2017,yang_stronger_2019}. This suggests that tuning of features in the phonon dispersion may alter the phonon scattering selection rules, potentially giving rise to a large modulation of $\kappa_{\rm L}$. In fact, in addition to the features described above, the phonon dispersion of BAs also displays a narrow optical phonon bandwidth \cite{ravichandran_non-monotonic_2019} and degeneracy between two TA branches and between two TO branches along high-symmetry paths, i.e., $ \Gamma$-$\rm{X}$ and $ \Gamma$-$\rm{L}$, as seen in Fig.~\ref{Fig2}(b). Given the difficulty in satisfying energy conservation, the degenerate TO branches restrict scattering channels involving two optical and one acoustic phonon (AOO), while the degenerate TA modes weaken AAA processes.

Symmetry analysis reveals that these degenerate phonon bands are protected by \({3}^{+}_{111}\) and \({m}_{110}\) symmetries, which can be broken by applying a uniaxial strain along the [100] crystal direction, as illustrated in Fig.~\ref{Fig2}(a). Taking 14\% tensile strain as a representative, it is evident from Fig.~\ref{Fig2}(c) that the presence of tensile strain leads to the lifting of TA and TO band degeneracies along the $\Gamma$-$ \rm{X}$ and $\Gamma$-$\rm{L}$ directions (parallel to the $b$-axis), and the degree of band splitting increases with the increase of strain (see Supplementary Material \cite{ckl_supplement}). Meanwhile, we notice that in the presence of strain, the overall phonon spectrum softens, especially for the optical branches, leading to a substantial increase in the phonon population. Additionally, it is found that a large A-O gap still persists even under the strain, suggesting that scattering channels involving AAO processes remain strongly suppressed. 

\begin{figure}[htp]
\centering
\includegraphics[width=\columnwidth]{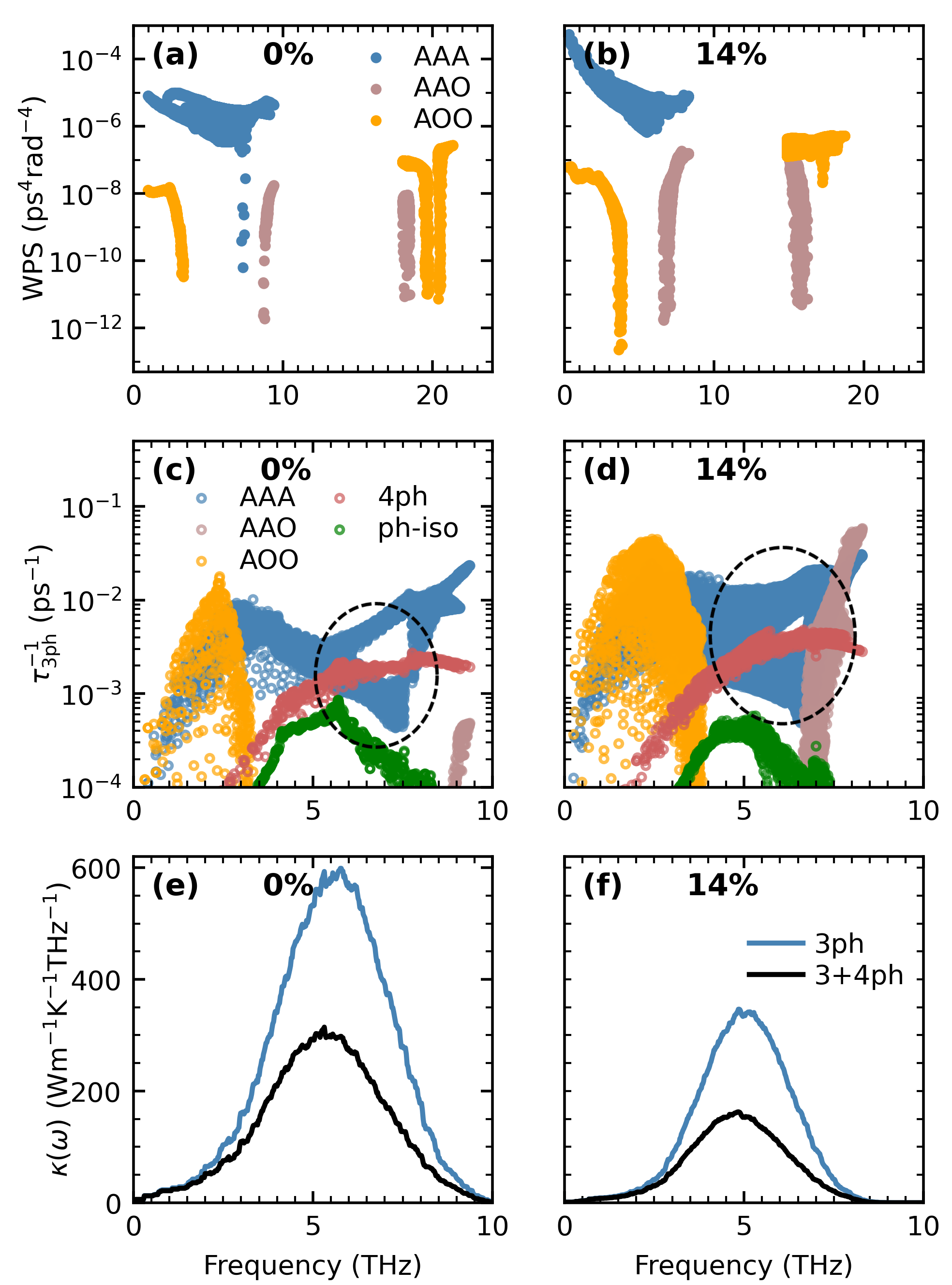}
\caption{Contributions to the WPS from different 3ph processes for unstrained (a) and 14\%-strained (b) BAs. (c, d) Sub-process classified 3ph, total 4ph, and ph-iso scattering rates of acoustic phonons at 300 K and different strains. (e, f) Spectral $\kappa_{\rm L}$ along the $b$ axis at different strains, where the results with and without including 4ph scattering are compared. }
\label{Fig3}
\end{figure}

From a vibrational point of view, the splitting of degenerate phonons facilitates the fulfillment of scattering selection rules for 3ph processes. This change, combined with the increase of phonon population resulting from the phonon softening, would significantly increase the number of 3ph scattering channels. To clarify this, we compare the mode projected 3ph scattering phase space for the intrinsic and strained BAs in Figs.~\ref{Fig2}(b) and (c). It is seen that under the action of strain, a pronounced enhancement in the 3ph phase space of the overall phonon branches is observed, particularly for the splitting TO and TA modes, as marked by the rectangles in Fig.~\ref{Fig2}(c).
          
To delve deeper into the impact of the phonon splitting on the 3ph scattering, we decompose the 3ph phase space into AAA, AOO, and AAO processes, as given in Figs.~\ref{Fig3}(a) and (b). It clearly shows that AAA channels predominantly contribute to the 3ph phase space for heat-carrying acoustic phonons ($<$10 THz). When compared to the intrinsic case, all scattering channels due to AAA, AOO, and AAO processes exhibit a substantial increase upon the application of strain. Although these processes display a similar response to strain, the underlying origins are different. Specifically, under the action of strain, the TA phonon splitting partially mitigates the bunching effect of acoustic branches on the AAA selection rule. As a result, more low-frequency acoustic phonons are allowed to participate in AAA scattering processes. For instance, the process TA + TA $\rightarrow$ TA/LA becomes more feasible by absorbing a lower-frequency TA phonon. By comparing Figs.~\ref{Fig3}(a) and (b), it is seen that the sharp dip in the AAA phase space, indicative of the bunching effect, is obviously weakened after the strain is applied. Besides, the TO phonon degeneracy lifting facilitates the participation of lower-frequency acoustic phonons in the AOO processes, i.e., the process TO + TA/LA $\rightarrow$ TO involving a low-frequency TA/LA mode is easier to occur. In addition, the increased optical bandwidth under strain also extends the frequency range of acoustic phonons involved in the AOO processes. Both factors contribute to the enlargement of the AOO phase space. In contrast, the enhancement of the AAO phase space is attributed to the slight narrowing of the A-O gap and is unrelated to the phonon splitting.

As is shown in Figs.~\ref{Fig3}(c) and (d), the enlarged phase space revealed above significantly increases the 3ph scattering rates of acoustic phonons ($<$ 10 THz), which contribute the most to the heat conduction as evidenced by the spectral $\kappa_{\rm L}$ in Figs.~\ref{Fig3}(e) and (f). Notably, as indicated by the ovals in Figs.~\ref{Fig3}(c) and (d), the dip in the 3ph scattering rates is markedly weakened due to the increased AAA channels when imposing the strain. As a result, the significant enhancement of 3ph scattering reduces the $b$-axis $\kappa_{\rm L}$ by 54\% at RT. On the other hand, we notice that 4ph scattering rates, which are governed by the AAOO processes \cite{feng_four-phonon_2017,yang_stronger_2019}, are unrestricted to the features of phonon dispersion but also increase considerably. This arises from the strain-induced phonon softening, which enhances the phonon population and, consequently, amplifies the availability of 4ph scattering channels (see Supplementary Material \cite{ckl_supplement}). Additionally, the isotope scattering is extremely weak regardless of the presence of strain, rendering its effect on $\kappa_{\rm L}$ negligible. These results indicate that uniaxial strains simultaneously intensify the 3ph and 4ph scattering, in stark contrast to the isotropic strain that drives the opposite response of 3ph and 4ph scattering \cite{ravichandran_non-monotonic_2019,li_anomalous_2022}. When accounting for 4ph scattering, the application of strain reduces the $b$-axis $\kappa_{\rm L}$ by 59\%. 

\subsection{Anomalous thermal anisotropy induced by uniaxial strain}
\begin{figure*}[htp]
\centering
 \includegraphics[width=2\columnwidth]{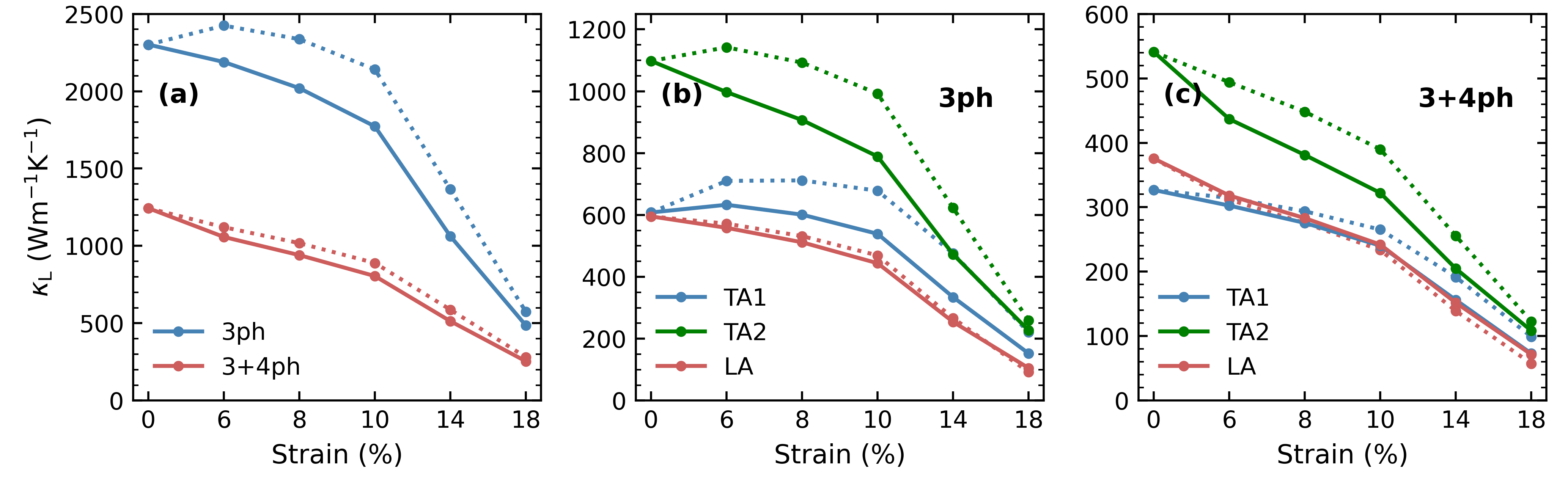}
\caption{(a) The uniaxial strain dependence of $\kappa_{\rm L}$ at RT. Contribution to $\kappa_{\rm L}$ from each acoustic branch without (b) and with (c) 4ph scattering included. The dotted and solid curves correspond to the $\kappa_{\rm L}$ along the $a$ and $b$ axes, respectively.}
\label{Fig4} 
\end{figure*}

Since uniaxial strain induces lattice anisotropy, the $\kappa_{\rm L}$ along different crystallographic directions ought to behave anisotropically. The strain-dependent $\kappa_{\rm L}$ at RT along the uniaxial stretching direction (corresponding to $a$-axis) and perpendicular to the strain direction (corresponding to $b$-axis) is plotted in Fig.~\ref{Fig4}(a). It can be seen that uniaxial strain suppresses the $b$-axis $\kappa_{\rm L}$ more pronouncedly than that along the $a$-axis, especially when only 3ph scattering is considered. At $\varepsilon$ = 6\%, the 3ph theory predicts a considerable reduction in the $b$-axis $\kappa_{\rm L}$, while exhibiting an abnormal increase in the $a$-axis $\kappa_{\rm L}$. After including 4ph scattering, the $\kappa_{\rm L}$ in both axes reduces by $\sim$15\%. Under the action of large strains, e.g., at $\varepsilon$ = 18\%, the 3ph-limited $\kappa_{\rm L}$ along the $b$-axis drops dramatically from 2302 to 485 W/mK at RT, by 79\%. When 4ph scattering is further included, the $b$-axis $\kappa_{\rm L}$ decreases from 1244 to 254 W/mK, a reduction of $\sim$80\%. As the strain continues to increase, the electronic band gap of BAs gradually narrows (see Supplementary Material \cite{ckl_supplement}), triggering a transition from a semiconductor to a metal. This scenario lies outside the scope of the current study.

Further decomposing $\kappa_{\rm L}$ into different branches in Figs.~\ref{Fig4}(b) and (c), we observe that the strain responses of different phonon branches to $\kappa_{\rm L}$ vary largely along different axes. At the 3ph scattering level, the contributions of both TA branches to the $a$-axis $\kappa_{\rm L}$ display weak strain sensitivity at strains below 10\%, with even noticeable increases observed at $\varepsilon$ = 6\%. This positive response directly explains the observed anomalous increase in the $a$-axis $\kappa_{\rm L}$. In contrast, along the $b$-axis, the TA branch contributions decrease significantly with increasing strain, particularly for the high-lying TA (denoted as TA2) branch. We also find that the LA branch exhibits nearly identical strain responses along both axes. These findings reveal that the anisotropic thermal response originates primarily from strain-induced lifting of TA phonon degeneracy, which effectively suppresses thermal transport parallel to the crystallographic $b$-axis due to the weakening of the acoustic phonon bunching effect.

However, the inclusion of 4ph scattering significantly alters this behavior, causing a monotonic reduction in the contributions to $\kappa_{\rm L}$ from all three branches with increasing strain for both axes, as seen in Fig.~\ref{Fig4}(c). It is important to note that although 4ph scattering partially reduces the $\kappa_{\rm L}$ anisotropy between the two axes, the strain-induced suppression of TA branch contributions remains significantly more pronounced along the $b$-axis than along the $a$-axis. This confirms that the anomalously larger reduction of $\kappa_{\rm L}$ along the $b$-axis than along the $a$-axis is dominated by strain-mediated 3ph processes, which are closely associated with the TA band degeneracy lifting.

\subsection{Concurrent enhancement of three- and four-phonon scattering}
\begin{figure}[htp]
\centering
\includegraphics[width=\columnwidth]{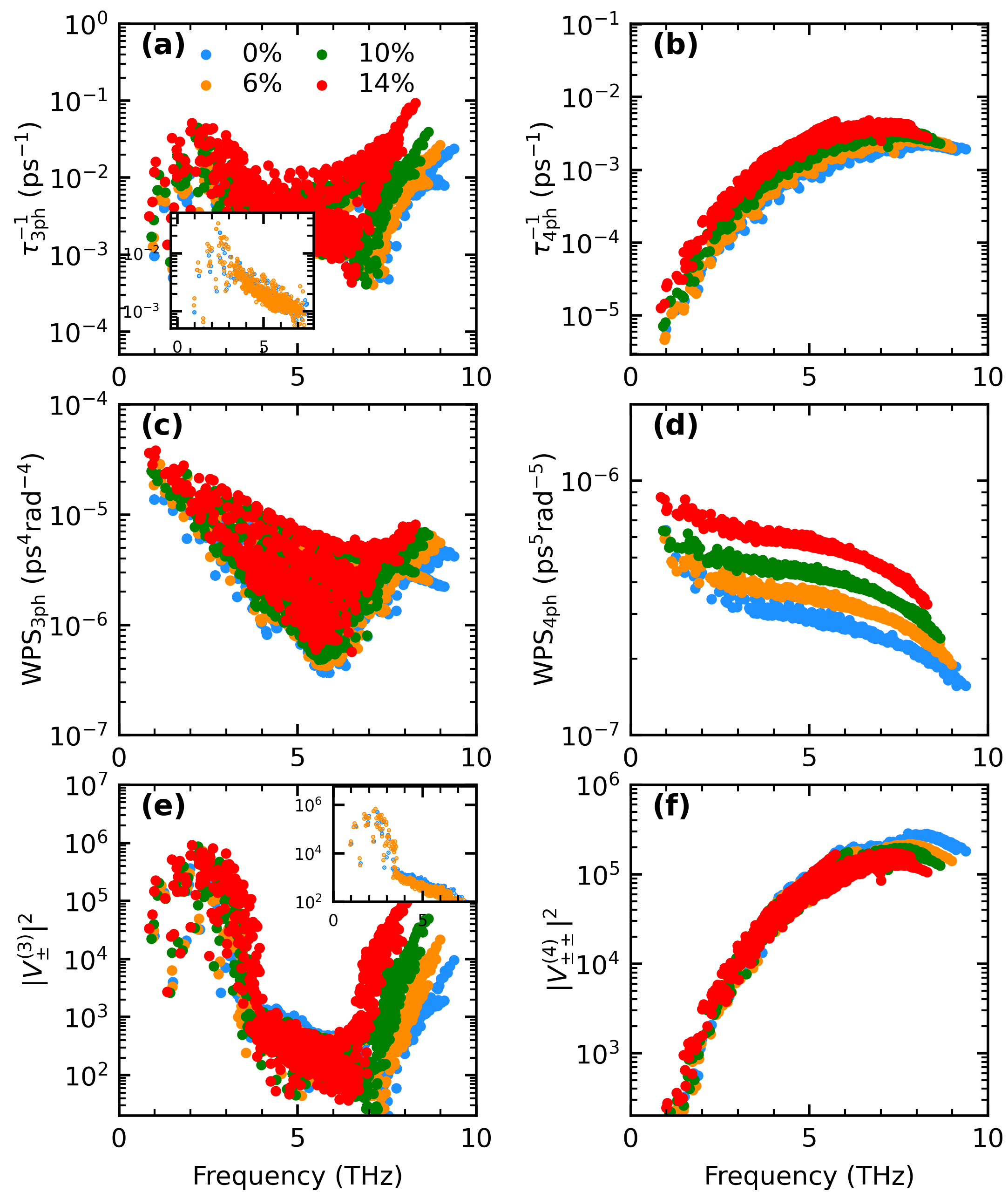}
\caption{The scattering rates for 3ph processes (a) and 4ph processes (b) under different uniaxial strains. (c, d) Corresponding WPS due to 3ph processes and 4ph processes. (e) 3ph scattering matrix elements ($|V_{ \pm}^{(3)}|^2$) with respect to frequency, with the inset illustrating the $|V_{ \pm}^{(3)}|^2$ for TA modes under 0\% and 6\% strains. (f) 4ph scattering matrix elements ($|V_{ \pm \pm}^{(4)}|^2$) under different strains. }
\label{Fig5}
\end{figure}

To elucidate the strain-dependent $\kappa_{\rm L}$ behavior, we present the calculated 3ph and 4ph scattering rates under various strains in Figs.~\ref{Fig5}(a) and (b). As clearly shown in the figures, the overall scattering rates for both 3ph and 4ph processes increase monotonically with strain, consistent with the monotonic reduction of $\kappa_{\rm L}$ as strain rises. It is noteworthy that 3ph scattering rates of TA phonons below 5 THz show a slight decrease under 6\% strain, as seen in the inset of Fig.~\ref{Fig5}(a), accounting for the aforementioned increase in the TA branch contribution to the 3ph-limited $\kappa_{\rm L}$. The influence of strain on scattering rates manifests in two aspects: its effect on scattering phase space and its impact on scattering matrix elements \cite{wang_synergistic_2021,yang_reduced_2022,wei_tensile_2024,ding2024anharmonicity}. To clarify which one is responsible for the intensified phonon scattering by strain, we detect these two factors for 3ph and 4ph processes. From Figs.\ref{Fig5}(c) and (d), it is shown that both the 3ph and 4ph phase spaces of heat-carrying acoustic phonons display a pronounced monotonic enhancement with increasing strain. Meanwhile, Figs.~\ref{Fig5}(e) and (f) reveal that $|V_{ \pm}^{(3)}|^2$ experience a substantial increase under strain, while $|V_{ \pm \pm}^{(4)}|^2$ for acoustic phonons even exhibit a slight decrease within the mid-to-high frequency range. This suggests that uniaxial strain significantly strengthens the third-order anharmonicity while slightly weakening the fourth-order anharmonicity. Based on the results, we conclude that the elevated 3ph scattering rates stem from the concurrent enhancement of 3ph phase space and third-order anharmonicity, while the increased 4ph scattering rates are fully governed by the expanded 4ph phase space. Additionally, as illustrated in the inset of Fig.~\ref{Fig5}(e), the $|V_{ \pm}^{(3)}|^2$ of TA phonons shows a slight decline below 5 THz under 6\% strain, suggesting that the increased TA scattering rates at this strain level arise from the weakened third-order anharmonicity in the low-frequency ($<$ 5 THz) range. A similar phenomenon was also reported in diamond \cite{wang2022anomalous}. 

%For the TA2 branch in particular, their contribution to $\kappa_{\rm L}$, which is dominated by AAA processes \cite{lindsay_first-principles_2013}, exhibits the most significant suppression due to the strain-enhanced AAA scattering. As evident in Fig.~\ref{Fig5}(b), the anisotropy in $\kappa_{\rm L}$ originates mainly from the contribution of TA2 phonons, which respond differently to strain along different paths as mentioned above. 

\begin{figure}[htp]
\centering
\includegraphics[width=\columnwidth]{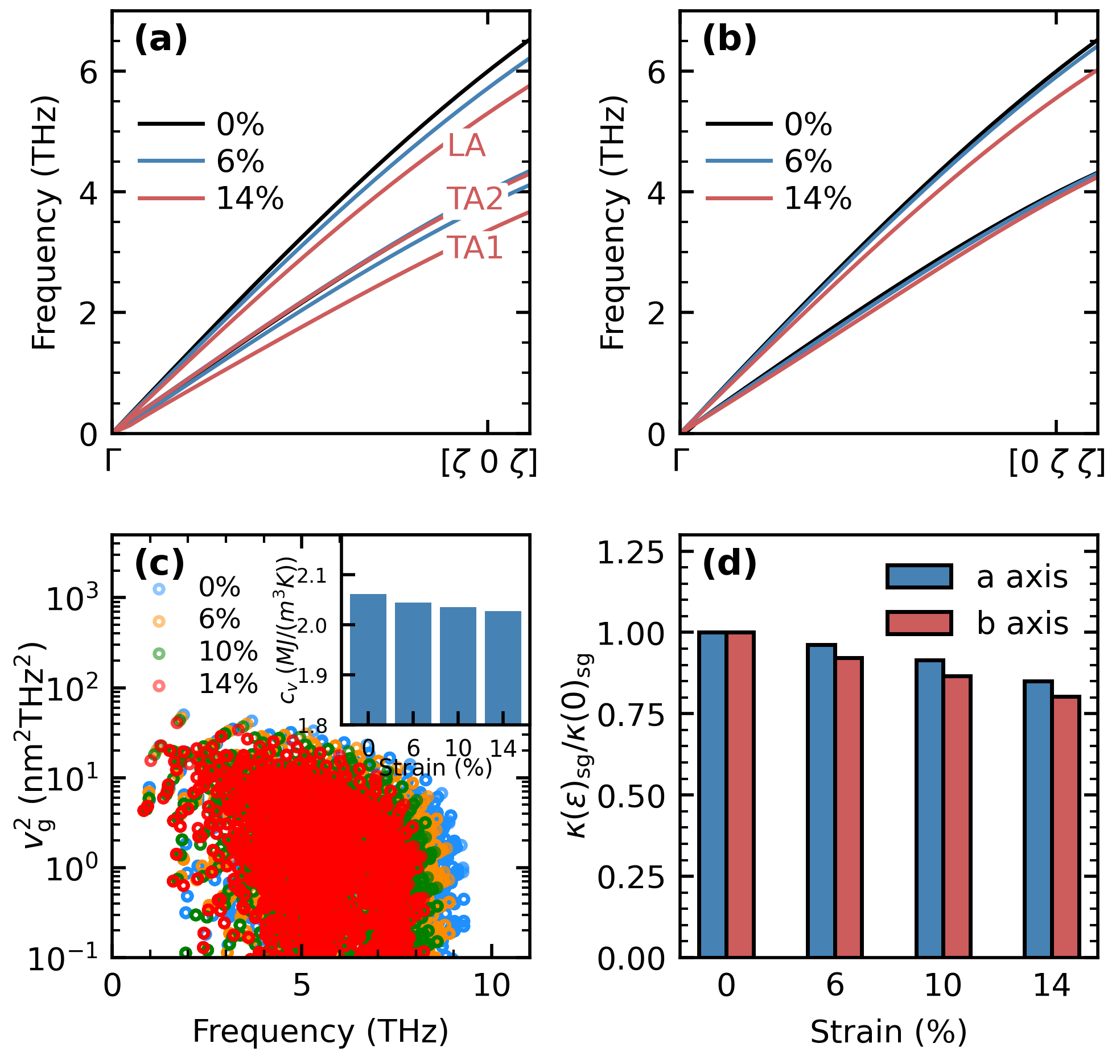}
\caption{(a, b) Strain-dependent acoustic phonon dispersions along different paths. (c) Squared group velocities along the $b$ axis under different strains. The inset shows the strain-dependent volumetric heat capacity at RT. (d) Normalized $\kappa_{\rm SG}$ along the $a$ and $b$ axes at different strains, respectively. }
\label{Fig6}
\end{figure}

In fact, strain modulates $\kappa_{\rm L}$ through multiple mechanisms: it not only influences phonon lifetimes via tuning phonon-phonon interactions but also affects phonon group velocities and heat capacity. From Figs.~\ref{Fig6}(a) and (b), it is seen that the LA branch monotonically decreases with increasing strain, while changes in the TA branches differ depending on the crystal direction. The phonon splitting occurs along the [$\zeta$ 0 $\zeta$] direction (which aligns with the $b$-axis). In this direction, the TA1 branch softens obviously, whereas the TA2 branch remains almost unchanged under strains, and the separation between the TA1 and TA2 branches widens as the strain increases. Along the [0 $\zeta$ $\zeta$] direction, corresponding to the $a$-axis, both TA branches exhibit a slight decrease when subjected to strain. These changes in the phonon spectrum lead to an obvious reduction in the group velocity of acoustic phonons, accompanied by minor variations in heat capacity, as clearly depicted in Fig.~\ref{Fig6}(c) and Fig. S6. Concurrently, these changes bring about differences in the group velocities of TA phonons along the $a$ and $b$ axes, partially accounting for the observed $\kappa_{\rm L}$ anisotropy. 

To quantify the impact of strain on the harmonic component of $\kappa_{\rm L}$, we calculate the small-grain-limit reduced thermal conductivity, defined as $\kappa_{\rm SG}=\sum_{p,\mathbf{q}}C_{V}(p\mathbf{q})v_{{p\mathbf{q}}}$ \cite{li_shengbte_2014}, at different strains. Fig.~\ref{Fig6}(d) shows the RT $\kappa_{\rm SG}$ along the $a$ and $b$ axes for different strains, scaled by the intrinsic values. It is observed that $\kappa_{\rm SG}$ decreases noticeably with increasing strain, showing a reduction of up to 24\%. This reduction is well below the actual decline in $\kappa_{\rm L}$, which reaches up to 80\%. This means that the strain-induced suppression of $\kappa_{\rm L}$ is mainly driven by the intensified phonon scattering, while the alterations in the harmonic component of $\kappa_{\rm L}$ play a secondary role.

\section{Conclusions}
In summary, we have elucidated the profound effect of symmetry-breaking uniaxial strain on lowering the thermal conductivity of BAs from first-principles. Our calculation reveals that uniaxial strain causes the $\kappa_{\rm L}$ of BAs to monotonically decrease as the strain increases, owing to the concurrent enhancement of 3ph and 4ph scattering. This arises largely from the lifting of phonon band degeneracy and the softening of phonon dispersion, which substantially increase phonon-phonon scattering channels, particularly processes involving the AAA scattering. Remarkably, uniaxial strain leads to a strong suppression of $\kappa_{\rm L}$ by nearly 80\% at RT. Moreover, we find that uniaxial strain enables a pronounced anisotropic suppression of $\kappa_{\rm L}$, with significantly stronger reduction in the direction perpendicular to the strain compared to the strain axis. These findings uncover the crucial role of uniaxial strain in tuning anharmonic phonon-phonon interactions and offer a new route to engineer heat conduction in solids via breaking crystal symmetry.

\section{acknowledgments}
This work is supported by the National Natural Science Foundation of China (Grant No. 12374038 and No. 12404045), Fundamental Research Funds for the Central Universities of China (Grant No. 2023CDJKYJH104), and Chongqing Natural Science Foundation (Grant No. CSTB2022NSCQ-MSX0834).

% \bibliography{common} 

\end{document}